\documentclass[draftcls,12pt,onecolumn]{IEEEtran}

\usepackage{cite}
\usepackage[pdftex]{graphicx}
\usepackage[cmex10]{amsmath}
\usepackage{amssymb}
\usepackage{amsfonts}

\usepackage{algorithm}
\usepackage{algorithmic}
\usepackage[caption=false,font=footnotesize]{subfig}
\usepackage{color}
\usepackage{array}
\usepackage{mdwmath}
\usepackage{mdwtab}
\usepackage{eqparbox}
\usepackage[colorlinks=true,urlcolor=black]{hyperref}
\usepackage{multirow}
\usepackage{bigstrut}
\usepackage{booktabs}
\usepackage{makecell}
\usepackage{color}
\usepackage{algorithm}
\usepackage{algorithmic}
\usepackage[caption=false,font=footnotesize]{subfig}
\usepackage{color}

\begin{document}

\title{An Amateur Drone Surveillance System\\ Based on Cognitive Internet of Things}
\author{ \IEEEauthorblockN{Guoru~Ding, Qihui Wu, Linyuan Zhang, Yun Lin,\\ Theodoros A. Tsiftsis, and Yu-Dong Yao}




\thanks{Guoru Ding is with the National Mobile Communications Research Laboratory, Southeast University, Nanjing 210018, China and also with the College of Communications Engineering, PLA University of Science and Technology, Nanjing 210007, China (email: dr.guoru.ding@ieee.org).}
\thanks{Qihui Wu is with the Department of Electronics and Information Engineering, Nanjing University of Aeronautics and Astronautics, Nanjing 210007, China (email: wuqihui2014@sina.com).}
\thanks{Linyuan Zhang is with the College of Communications Engineering, PLA University of Science and Technology, Nanjing 210007, China (email: zhanglinyuan5@163.com).}
\thanks{Yun Lin is with the College of Information and Communication Engineering, Harbin Engineering University, Harbin, China (e-mail: linyun@hrbeu.edu.cn).}
\thanks{Theodoros A. Tsiftsis is with the School of Engineering, Nazarbayev University, Astana 010000, Kazakhstan (email: theodoros.tsiftsis@nu.edu.kz).}
\thanks{Yu-Dong Yao is with the Department of Electrical and Computer Engineering, Stevens Institute of Technology, Hoboken, NJ 07030, USA (e-mail: yyao@stevens.edu).}}
\maketitle

\begin{abstract}
Drones, also known as mini-unmanned aerial vehicles, have attracted increasing attention due to their boundless applications in communications, photography, agriculture, surveillance and numerous public services. However, the deployment of amateur drones poses various safety, security and privacy threats. To cope with these challenges, amateur drone surveillance becomes a very important but largely unexplored topic. In this article, we firstly present a brief survey to show the state-of-the-art studies on amateur drone surveillance. Then, we propose a vision, named \emph{Dragnet}, by tailoring the recent emerging cognitive internet of things framework for amateur drone surveillance. Next, we discuss the key enabling techniques for Dragnet in details, accompanied with the technical challenges and open issues. Furthermore, we provide an exemplary case study on the detection and classification of authorized and unauthorized amateur drones, where, for example, an important event is being held and only authorized drones are allowed to fly over.
\end{abstract}

\begin{IEEEkeywords}
Amateur drone surveillance, Cognitive internet of things, Unmanned aerial vehicles, Anti-drone technology
\end{IEEEkeywords}

\IEEEpeerreviewmaketitle


\section{Introduction}

During the past few years, mini-unmanned aerial vehicles (UAV), also known as drones, have received worldwide increasing interests for their numerous applications in communications, surveillance, agriculture, photography, and public services, etc \cite{IoT-survey}. For instance, the registered number of drones in use in the U.S. exceed 200,000 just in the first 20 days of January 2016, after the Federal Aviation Administration (FAA) started requiring owners to sign up. However, the deployment of drones, especially the amateur drones, poses critical challenges. First of all, drones that enter the airspace around airports can pose serious safety threats to the conventional air traffic via physical collisions or wireless interference. Similarly, they would also bring various public risks to national institutions and assets (e.g., nuclear power plants, historical sites, and governmental houses), because of their ability to carry the explosive and other destructive chemicals or agents. Moreover, amateur drones are increasingly being used for spying/reconnaissance applications that violate military secrets and personal privacy, e.g. by taking photos/videos without permission. In addition, drones may cause physical harms to people on the ground.

In order to cope with these challenges, amateur drone surveillance becomes a very important but largely unexplored area. As one of the well-known efforts, the FAA published more than 200 facility maps in 27 April 2017 to streamline the commercial drone authorization process, where the maps depict areas and altitudes near airports that drones may operate safely \cite{FAA-map}. To keep watch on potential misbehaved drones, the FAA is coordinating with government and industry partners to evaluate technologies that could be used to detect drones in and around airports. In parallel, regulatory activities for drones have also started around the world, such as the regulation proposal from the European Aviation Safety Agency (EASA), which defines the technical and operational requirements for the drones. Technical requirements refer for example to the remote identification of drones while operational requirements refer to a system that ensures the drones does not fly above a prohibited zone \cite{EU-EASA}.

In this article, we firstly present a brief survey on the state-of-the-art studies on amateur drone surveillance, where the existing anti-drone technologies and the well-known anti-drone systems are discussed. Then, we propose a vision named \emph{Dragnet}\footnote{In this article, \emph{Dragnet} takes its name from a famous American radio, television and motion picture series, enacting the cases of a dedicated Los Angeles police detective, Sergeant Joe Friday, and his partners, meaning a system of coordinated measures for apprehending criminals or suspects, with a by-product to earn praise for improving the public opinion of drone surveillance.} by tailoring the recent emerging cognitive internet of things framework for amateur drone surveillance. Next, we discuss the key enabling techniques accompanied with the technical challenges and open issues ahead. Furthermore, we provide an exemplary case study on the detection and classification of authorized and unauthorized amateur drones from a multi-hypothesis testing perspective.

\section{Amateur Drone Surveillance: State-of-the-Art}

It is well known that military UAVs have been used in battlefield for several decades, however, the boundless applications of civilian or commercial drones have received worldwide interests just during the past few years. Consequently, amateur drone surveillance is a relatively new and largely unexplored area. Intuitively, one can take lessons from military anti-drone technologies, where majority of the existing related studies can be generally grouped into four classes:
\begin{itemize}
  \item \emph{Warning technique}, where various detection devices such as ground cameras, sensors and radars are deployed to perform early warning on the presence of any drone \cite{Counter-measures}.
  \item \emph{Spoofing technique}, where electronic, optical and/or infrared technologies are used to send false signals for abducting drones to land, with the GPS spoofing as a typical example \cite{Spoofing}.
  \item \emph{Jamming technique}, where the control and navigation information is disturbed by posing strong interference via for example electromagnetic gun, which makes the automatic driving system and/or the communication system of drones become invalid \cite{jamming}.
  \item \emph{Mitigation technique}, where drones can be destroyed or captured, mainly for emergency, by micro-missile, laser or arrest net, etc \cite{mitigation}.
\end{itemize}

More interestingly, there are already several anti-drone systems that utilize the above technologies. To facilitate general readers, in Table I, we present a brief summary of several well-known anti-drone systems, where the key functionalities and the useful link for each system are list for reference. Notably, we do not try to present a comprehensive survey here, it is expected that many more anti-drone systems are being developed all over the world.

\begin{table*}[!t]
\caption{A brief summary of several well-known anti-drone systems.}
  \centering
    \begin{tabular}{|p{0.01\textwidth}|p{0.1\textwidth}|p{0.3\textwidth}|p{0.3\textwidth}|}
    \hline
    \multicolumn{2}{|c|}{\textbf{Systems}} & \textbf{Functions} & \textbf{Advantages and Limitations}\\\hline
    \multicolumn{2}{|c|}{\makecell[l]{\textbf{Anti-UAV Defense System (AUDS)}\\(URL: http://www.blighter.com/products/\\auds-anti-uav-defence-system.html)} } &AUDS is a smart-sensor and effector package capable of remotely detecting small UAVs and then tracking and classifying them before providing the option to disrupt their activity.
    &1) It has an electronic-scanning radar for target detection, an electro-optical vedio for tracking and classification, and a software defined intelligent directional RF inhibitor; 2) detection range: 10 km; 3) minimum target size: 0.01 m$^2$; 4) works in various weather conditions, 24 hours a day.\\\hline
    \multicolumn{2}{|c|}{\makecell[l]{\textbf{Dedrone: Automatic Anti-Drone Security}\\(URL: https://www.dedrone.com/en/)}}
    & Dedrone is an airspace security platform that detects, classifies, and mitigates all drone threats.
& 1) It enables airspace surveillance 24/7; 2) automated alarm $\&$ notification; 3) multi-sensor analysis: RF sensor and vedio sensor; 4) DroneDNA analysis and pattern recognition capabilities for drone classification.\\\hline
    \multicolumn{2}{|c|}{\makecell[l]{\textbf{The US Army's Enhanced Area Protection}\\ \textbf{and Survivability (EAPS) Technology}\\(URL: http://www.army-technology.com/news/\\newsus-army-engineers-demonstrate-eaps\\-capability-to-counter-uas-threats-4688379)}}
    & The EAPS is a missile-based counter rocket, artillery, and mortar defence system, which has been expanded to include threats from unmanned aircraft systems or drones.
    &1) It uses a 50mm cannon to launch command guided interceptors, and uses a precision tracking radar interferometer as a sensor, a fire control computer, and a radio frequency transmitter and receiver to launch the projectile into an engagement 'basket';  2) computations are done on the ground, and the radio frequency sends the information up to the round. \\\hline
    \multicolumn{2}{|c|}{\makecell[l]{\textbf{Boeing's Compact Laser Weapons System (CLWS)}\\ \textbf{Tracks and Disables UAVs}\\(URL: http://www.boeing.com/features/2015/\\08/bds-compact-laser-08-15.page)}} &CLWS is a laser weapon system that can be used to acquire, track, and identify a target, or even destroy it. & 1) It is portable and can be assembled in 15 minutes; 2) destroy targets such as UAVs from 22 miles away, 20 pounds with a 10-second with an energy beam of up to 10 kilowatts. \\\hline
\end{tabular}%
\label{tab:algorithms}
\end{table*}

Although the existing anti-drone technologies and systems can serve a good starting point for the research on amateur drone surveillance, there are still many issues open for solutions. First of all, amateur drones are generally small in size, light in weight and they appear in sudden, which make them hard to be detected. Second, the integration of artificial intelligence and advanced materials into the design of future amateur drones will significantly improve their capability in counter-capturing and stealth. Third, most of the existing anti-drone technologies are designed for specific drones or scenarios via relatively few technical methods, which limit their applications and generalization. The corresponding anti-drone systems are designed in isolation for specific military or civilian purposes, which are lack of networked information processing capability.

\section{Dragnet: Cognitive Internet of Things-Enabled Amateur Drone Surveillance}
Majority of the related work have focused on how to enable various individual surveillance devices or systems to see, hear and sense the physical world for drone surveillance. Making surveillance devices or systems connected to share the observations and to accomplish information fusion represent a research trend. In this article, we argue that only connected is not enough, beyond that, surveillance devices should have the capability to learn, think, and understand both physical and social worlds by themselves. This practical need impels us to develop a new vision, named \emph{Dragnet}, i.e., cognitive Internet of Things-enabled amateur drone surveillance, in order to empower the amateur drone surveillance with a ``brain'' for high-level intelligence\footnote{In this paper, the terms ``brain", ``intelligence", ``learning" and ``global decision" are mentioned. As the name implies, they are closely related with the hot fields of artificial intelligence and machine learning, where various mathematic tools (such as statistics, optimization, game theory, etc) have been invoked to develop powerful algorithms, with Alphago known as a recent big success to challenge and defeat the human go champion. Similarly, there is a clear trend to bridge the research between wireless communications and artificial intelligence.}. Before going deep into the new vision Dragnet and its enabling techniques, let's first share a brief background on cognitive internet of things to facilitate general readers.

\subsection{Cognitive Internet of Things}
The Internet of Things (IoT) is a technological revolution that brings us into a new ubiquitous connectivity, computing, and communication era. In the past decade, we have witnessed worldwide efforts on the research and development of IoT from academic community, service providers, network operators, and standard development organizations, etc. Technically, most of the attention has been focused on aspects such as communication, computing, control, and connectivity \cite{CRIoT}, which are indeed very important topics. However, without comprehensive cognitive capability, IoT is just like an awkward stegosaurus: all brawn and no brains. To fulfill its potential and deal with growing challenges, in our previous work~\cite{CIoT}, we propose to take the cognitive capability into consideration and empower IoT with high-level intelligence, and develop an enhanced IoT paradigm, i.e., brain-empowered IoT or cognitive IoT as follows~\cite{CIoT}:

\emph{``Cognitive IoT is a new network paradigm, where (physical/virtual) things or objects are interconnected and behave as agents, with minimum human intervention, the things interact with each other following a context-aware perception-action cycle, use the methodology of understanding-by-building to learn from both the physical environment and social networks, store the learned semantic and/or knowledge in kinds of databases, and adapt themselves to changes or uncertainties via resource-efficient decision-making mechanisms, with two primary objectives in mind:}
 \begin{itemize}
   \item \emph{bridging the physical world (with objects, resources, etc.) and the social world (with human demand, social behavior, etc.), together with themselves to form an intelligent physical-cyber-social (iPCS) system;}
   \item \emph{enabling smart resource allocation, automatic network operation, and intelligent service provisioning.''}
 \end{itemize}

\subsection{The Vision: Dragnet}
Now, we introduce a new vision for amateur drone surveillance, named \emph{Dragnet}, in order to empower the amateur drone surveillance with a ``brain'' for high-level intelligence by tailoring the recent advances in cognitive IoT. Here \emph{Dragnet} refers to a system of coordinated measures for apprehending criminals or suspects, with a by-product to earn praise for improving the public opinion on regulation-obeyed amateur drone operations.

\begin{figure*}[!t]
\centering
\includegraphics[width=.8\linewidth]{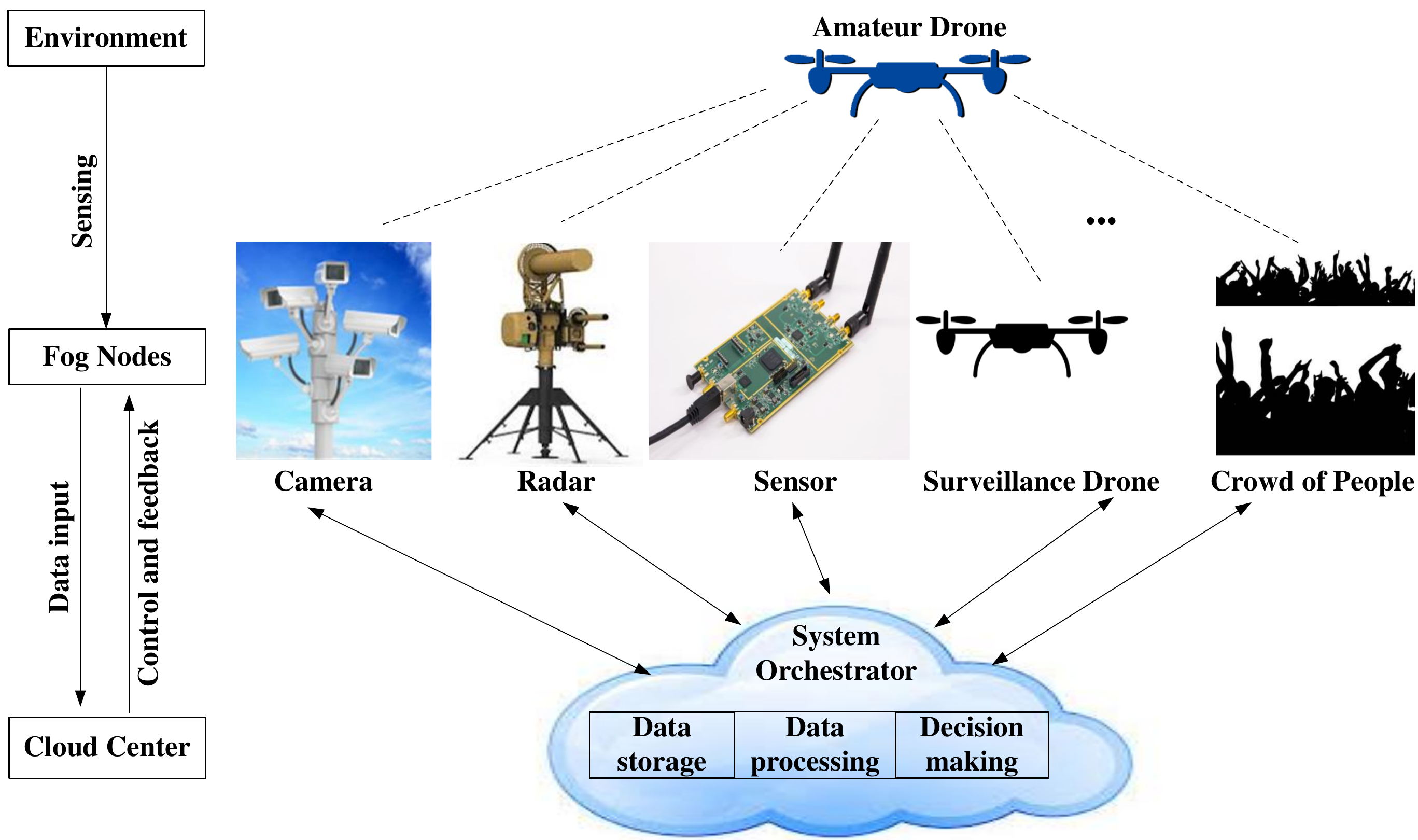}
\caption{Illustration of a joint fog-to-cloud computing framework for Dragnet-based amateur drone surveillance.}
\label{Fig-Dragnet}
\end{figure*}

As shown in Fig. \ref{Fig-Dragnet}, in the proposed Dragnet-based amateur drone surveillance, there is a joint fog-to-cloud computing hardware framework. Specifically, various active and passive surveillance devices (such as cameras, sensors, radars, and drones) or crowd of people serve as local fog computing\footnote{According to \cite{fog}, fog computing is a term for an alternative to cloud computing that puts a substantial amount of storage, communication, control, configuration, measurement and management at the edge of a network, rather than establishing channels for the centralized cloud storage and utilization, which extends the traditional cloud computing paradigm to the network edge.} platforms to sense the environment and locally warning the presence of amateur drone, while a cloud center acts as a system orchestrator that integrates the data from various fog nodes, stores and analyzes these data, and makes global decision-making on the presence of amateur drone as well as actions (e.g., jamming, capturing or destroy) on them. Notably, for the problem of amateur drone surveillance, we declare that both machine and human crowd intelligence can be employed to achieve a hybrid surveillance diversity and to enable a full-time-full-space surveillance. Generally, machines (e.g., cameras, sensors, radars) have relatively high surveillance accuracy but the number of them is limited. A large crowd of people can contribute surveillance data in the way, for example, mobile crowd sensing \cite{Mobile-crowdsensing} where individuals with sensing and computing devices (such as smartphones, in-vehicle sensors) collectively share data and extract information to measure and map phenomena on the presence of any amateur drone. The unique features of mobile crowd sensing include (but not limited to): i) large population of mobile devices bring massive surveillance data, especially for large-scale event (e.g., hot sports event) and activities (e.g., festival rally), where each person has one or more mobile devices to acquire and report data; ii) mobility of mobile devices enable ubiquitous surveillance coverage\footnote{It is an interesting issue on the design of incentive mechanisms for mobile crowd sensing. Similar issues have been extensively studied in the literature and majority of the existing mechanisms can be broadly classified into four groups: volunteer-based, interests-based, honor-based, and money-based. For the problem of interests in this paper, the governmental agency can probably distribute an App and provide some honors or small money to encourage people to share their surveillance data.}.

\begin{figure*}[!t]
\centering
\includegraphics[width=.8\linewidth]{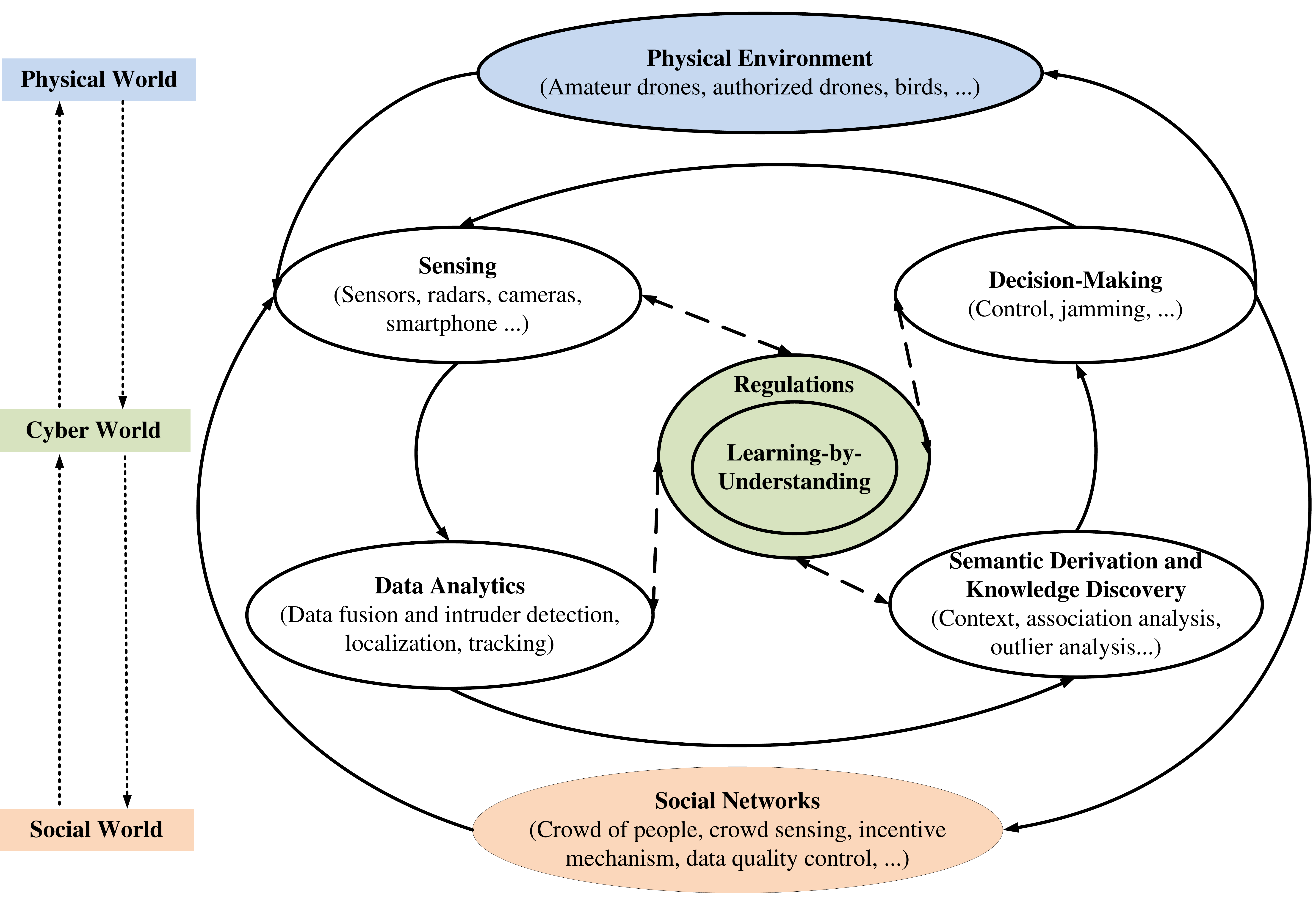}
\caption{Functional diagram of Dragnet-based amateur drone surveillance.}
\label{Fig2-CIoT-UAV}
\end{figure*}

Fig. \ref{Fig2-CIoT-UAV} presents the functional diagram of Dragnet-based amateur drone surveillance. Generally, Dragnet serves as a transparent bridge between physical world (with general physical/virtual things, amateur drones, authorized drones, birds, etc.) and social world (with human demand, social behavior, etc.), together with itself form an intelligent amateur drone surveillance system. With a synthetic methodology learning-by-understanding located at the heart, Dragnet consists of four key fundamental cognitive tasks, sequentially: 1) sensing; 2) data analytics; 3) semantic derivation and knowledge discovery; and 4) intelligent decision-making. Briefly, sensing is the most primitive cognitive task in Dragnet which serves as the input both from the physical environment and the social networks via various active and passive surveillance devices (such as cameras, sensors, radars) or crowd of people. Data analytics is one core cognitive task that performs intruder detection, localization, and tracking of amateur drones via mining kinds of surveillance data. Semantic derivation and knowledge discovery is the cognitive task that aims to make the objects in Dragnet self-understand and be aware, and to enable them to automatically derive the semantic from analyzed data, besides, based on the analyzed data and semantic, some valuable patterns or rules can be discovered as knowledge as well, which is a necessity for objects in Dragnet to be intelligent. Finally, decision-making is another core cognitive task that makes global decision on the presence of amateur drones as well as actions (e.g., jamming, capturing or destroy) to control them.
\begin{figure*}[!t]
\centering
\label{fig:Fig3-Technologies}
\includegraphics[width=0.8\linewidth]{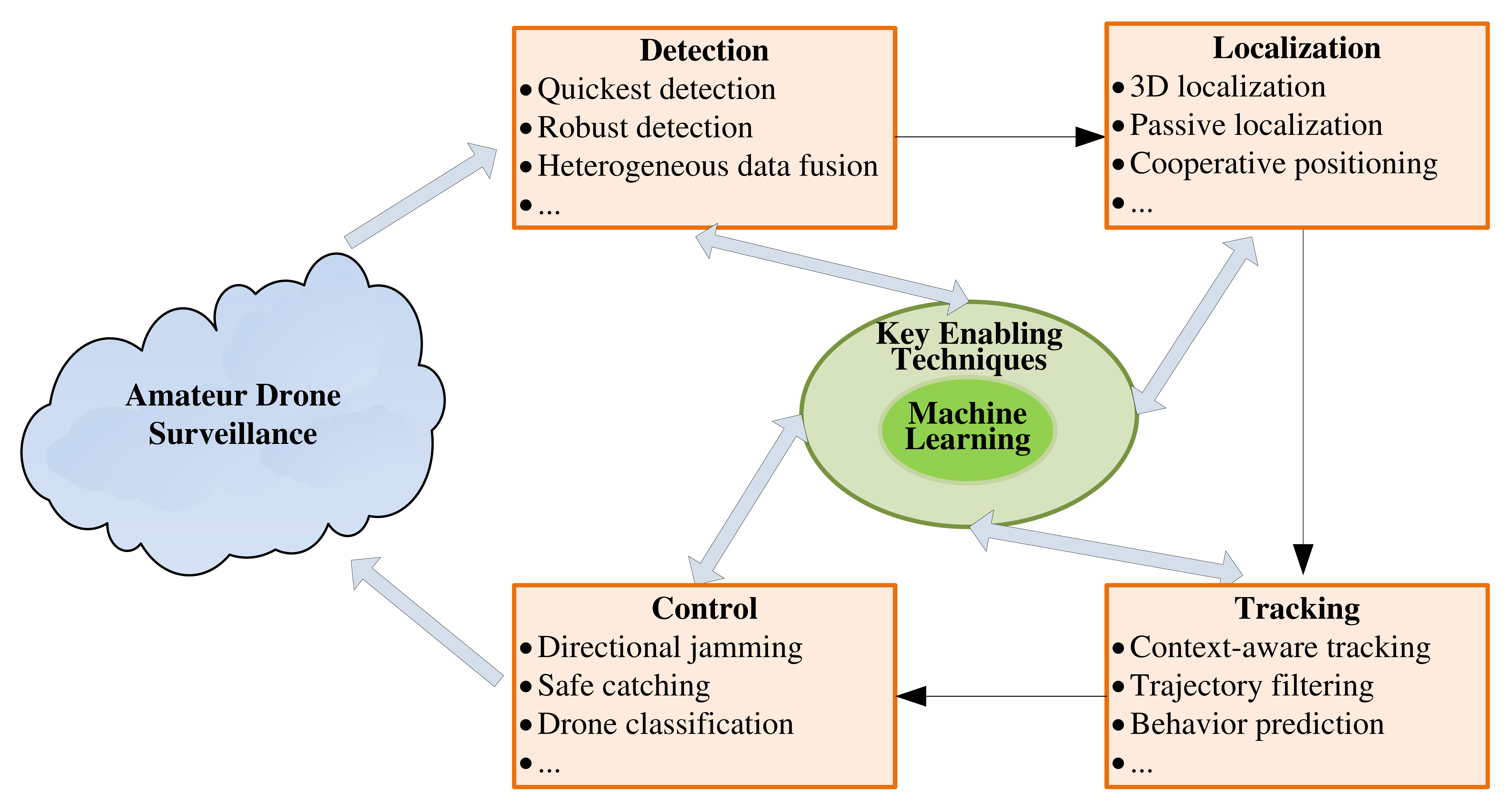}
\caption{Key enabling techniques for amateur drone surveillance.}
\label{Fig3-Technologies}
\end{figure*}

\section{Key Enabling Technologies for Dragnet}

To enable the cognitive tasks in Dragnet, there are many techniques. As shown in Fig. \ref{Fig3-Technologies}, among many others, the key enabling technologies for Dragnet mainly include: detection, localization, tracking, and control. In this section, we will present an overview on each of the key enabling techniques accompanied with the technical challenges and open issues ahead.

\subsection{Detection}
Considering the fact that amateur drones generally fly with low-altitude, there are a number of different approaches to drone detection: audio detection, video detection, thermal detection, radar detection, radio-frequency (RF) detection, etc \cite{Counter-measures}. However, each approach has its merits and limitations. For instance, noise from spinning propellers and electric motors can be detected by acoustic sensors such as microphones, but it can only detect the presence of a drone in the vicinity. Video cameras and thermal detection also provide a relatively short range detection but are subjected to weather conditions. Radars are well recognized as dominating drone detectors, but they also have difficulties in differentiating between those of a small drone and those from small birds and other sources of clutter. Moreover, radar detection has the requirement of continuous transmission of RF radiation. In an urban environment setting, placing a radar transmitter close to where people work and live may raise radiation health concerns. Using RF signals emitted by the drone is seen as another effective way to detect drones, while a drawback is that it must rely on the targets to transmit RF emissions\footnote{As a byproduct, RF signal-based detection can check the spectrum occupancy status of primary users with licensed spectrum bands and find white spaces for secondary communications systems without licensed bands \cite{1417,White-space}.}.

\subsubsection{Heterogeneous Data Fusion}
In order to improve the drone detection performance, it's a natural to combine the capability of different detection approaches for achieving a diversity gain. However, one challenge is to perform data fusion with heterogeneous sensors. Mathematically, random variables that characterize the data from heterogeneous sensors may follow disparate probability distributions. Performing optimal data fusion needs the joint probability density function of the heterogeneous data set. One often chooses to assume simple models such as the product model or multivariate Gaussian model, which lead to suboptimal solutions. Copula theory shows another approach to tackle heterogeneous data fusion, where the copulas function couples multivariate joint distributions to their marginal distribution functions \textcolor[rgb]{0.00,0.00,1.00}{\textbf{\cite{copula}}}.

\subsubsection{Quickest Detection}
The accuracy of detection is a basic requirement of the amateur drone surveillance system. However, the accurate detection is generally time-consuming. Moreover, amateur drones generally appear in sudden and can bring public risks in a very short time, e.g., impose explosive and other destructive chemicals to national institutions and assets. Thus, quickest detection is of great interest. As the name implies, quickest detection refers to real-time detection of any intruder amateur drone as quickly as possible, which can be casted in the framework of optimal stopping theory.


\subsection{Localization}
The next major step after detecting the amateur drone is the localization of the amateur drone intruder, which corresponds to a parameter estimation process and serves as the basis of the subsequential tracking and control operations. There are several unique features for the localization of the amateur drone intruder. The first one is that 3D position estimation algorithms are desired to accurately determine the position of the amateur drone (including latitude, longitude, and elevation), while majority of the state-of-the-art studies on localization focused on 2D position estimation. The second on is that passive localization algorithms are desired since the amateur drone intruder would not actively share its position or signal feature with the amateur drone surveillance system, which are quite different with the well-known GPS-based positioning and the Bluetooth/WiFi/cellular network-based localization. Actually, the surveillance system should not need the amateur drone to carry any responding/positioning equipment onboard.

\subsection{Tracking}
Drone localization and tracking have the inseparable connection. If localization is to obtain a location coordinate, tracking is to obtain a sequence of location coordinates. On the one hand, localization serves as the input and the basis for tracking. On the other hand, tracking can provide localization with prior information to correct the error of localization to some extent. For the tracking of the amateur drone, there are two vital techniques. One is context-aware tracking via trajectory filtering for moving/flying drones, the other is drone behavior prediction via mining of massive historical data on amateur drone activities.

For the surveillance of an amateur drone flying high, radar, video, and/or radio-frequency signal processing can be used to perform the localization and tracking (see, e.g., \cite{Counter-measures}). As shown in Table I, it is reported that in the well-known AUDS, it has an electronic-scanning radar for target detection of a range 10 km and an electro-optical video for tracking and classification of minimum target size 0.01 m$^2$. In practice, each approach has its merits and limitations. In order to improve the drone detection performance, it's a natural to combine the capability of different techniques for achieving a diversity gain.




\subsection{Control}
When an amateur drone intruder is detected and its location and flying trajectory is tracked, the last important step is to properly control the amateur drone. There are several popular drone control approaches: i) directional jamming the remote control signal or the navigation signal, ii) safe catching/hunting the low-altitude drone by a fishing net, iii) destroy the drone of high-threats via high-power laser, electromagnetic gun or micro-missile. Each approach has its application cases. Sometimes, we should make a choice, while other times we should adopt a combination. Currently, the drone control configuration of most drone surveillance systems are based on human operations. It's interesting and also challenging to exploit hybrid human and machine intelligence for highly-efficient drone control.

\subsection{Other Key Enabling Technologies}
Besides, there are several other key enabling technologies for Dragnet such as energy harvesting, full duplex, cognitive radio networks, etc. Specifically, energy harvesting and transfer is a promising technology to prolong the network lifetime where, traditionally, batteries are the primary energy source and regularly recharging or replacement of batteries are costly and inconvenient. Full duplex and cognitive radio networks can be used to improve the spectrum efficiency and spectrum utilization, respectively~\cite{full-duplex,1417}, especially when spectrum shortage occurs or licensed spectrum is unavailable.




\section{Case Study: Drone Detection and Classification}


Due to the page limit, here we only consider an example drone surveillance scenario that there may be unauthorized amateur drones intruding a certain protected region, where, for example, an important event is being held and only authorized drones are allowed to fly over. To enable the drone surveillance, we propose a ternary hypothesis testing as follows: $\mathcal{H}_0$, $\mathcal{H}_1$ and $\mathcal{H}_2$ denote the case that no drone exists, an authorized drone exists, and an unauthorized amateur drone exists, respectively. Then, the problem of interest is drone detection and classification, i.e., to detect whether a drone exists ($\mathcal{H}_1$/$\mathcal{H}_2$) or not ($\mathcal{H}_0$), and if there is a drone, to identify or classify whether it is an authorized drone ($\mathcal{H}_1$) or an unauthorized amateur drone ($\mathcal{H}_2$). Mathematically, we formulate an optimization problem as follows:
\begin{equation}
\begin{gathered}
  \max  \Pr ({\mathcal{H}_2}|{\mathcal{H}_2}), \hfill \\
  \text{subject~to}~~1-\Pr ({\mathcal{H}_0}|{\mathcal{H}_0}) \leq \alpha,~~1-\Pr ({\mathcal{H}_1}|{\mathcal{H}_1}) \leq \beta,  \hfill \\
\end{gathered}
\label{Eq_decision}
\end{equation}
where $\Pr ({\mathcal{H}_i}|{\mathcal{H}_i}),~i\in \{0,1\}$ is the probability of ${\mathcal{H}_i}$ being correctly detected. As shown in (\ref{Eq_decision}), the objective is to maximize the probability of correctly detecting the presence of the unauthorized amateur drone $\Pr ({\mathcal{H}_2}|{\mathcal{H}_2})$, while the constraints are set to control the false alarms. Specifically, the false alarm probability $1-\Pr ({\mathcal{H}_0}|{\mathcal{H}_0})=\Pr ({\mathcal{H}_1}|{\mathcal{H}_0})+\Pr ({\mathcal{H}_2}|{\mathcal{H}_0})$ corresponds to the case that an authorized or unauthorized drone is falsely detected while the truth is no drone being present. Similarly, the false alarm probability $1-\Pr ({\mathcal{H}_1}|{\mathcal{H}_1})=\Pr({\mathcal{H}_2}|{\mathcal{H}_1})+\Pr({\mathcal{H}_0}|{\mathcal{H}_1})$ corresponds to the false alarm that an unauthorized drone is falsely detected or no drone is detected while the truth is authorized being present.

\begin{figure*}[!t]
\centering
\includegraphics[width=0.5\linewidth]{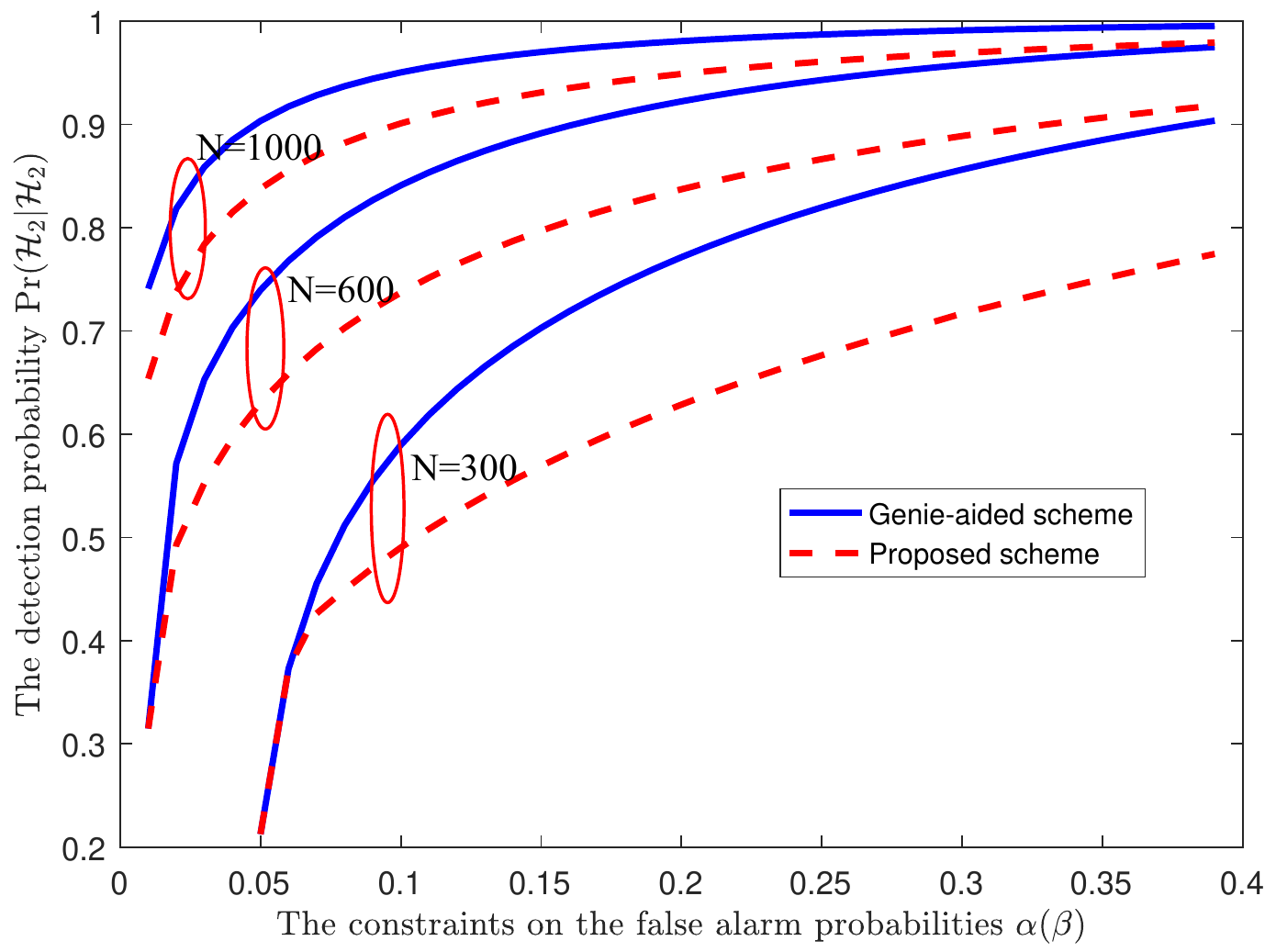}
\caption{Tradeoff between the detection probability and the constraints on the false alarm probabilities ($\alpha=\beta$).}
\label{Fig-roc-local}
\end{figure*}

Generally, we hope that the detection probability $\Pr ({\mathcal{H}_2}|{\mathcal{H}_2})$ is as large as possible, while the two false alarm probabilities are as small as possible. However, there is a tradeoff between them, just as shown in the simulation results of Fig. \ref{Fig-roc-local}, the probability of the unauthorized drone being correctly detected increases always at the price of an increase of the false alarms $\alpha$ and $\beta$, and vice versa. Specifically, in this figure, we propose a generalized likelihood ratio test scheme (named ``Proposed scheme'' for short in Fig. \ref{Fig-roc-local}) to present a solution to the optimization problem in (\ref{Eq_decision}), which considers the fact that the unknown characteristics of the unauthorized drones makes $\mathcal{H}_2$ a composite hypothesis. For the purpose of performance comparison, we also present a genie-adied scheme where both the distributions of the received signal strength from the authorized drone and the unauthorized drones are known as \emph{a prior}. We can see that the detection performance of both schemes can be improved by increasing the number of samples $N$, which corresponds to an increasing of the detection delay.

\begin{figure*}[!t]
\centering
\includegraphics[width=0.6\linewidth]{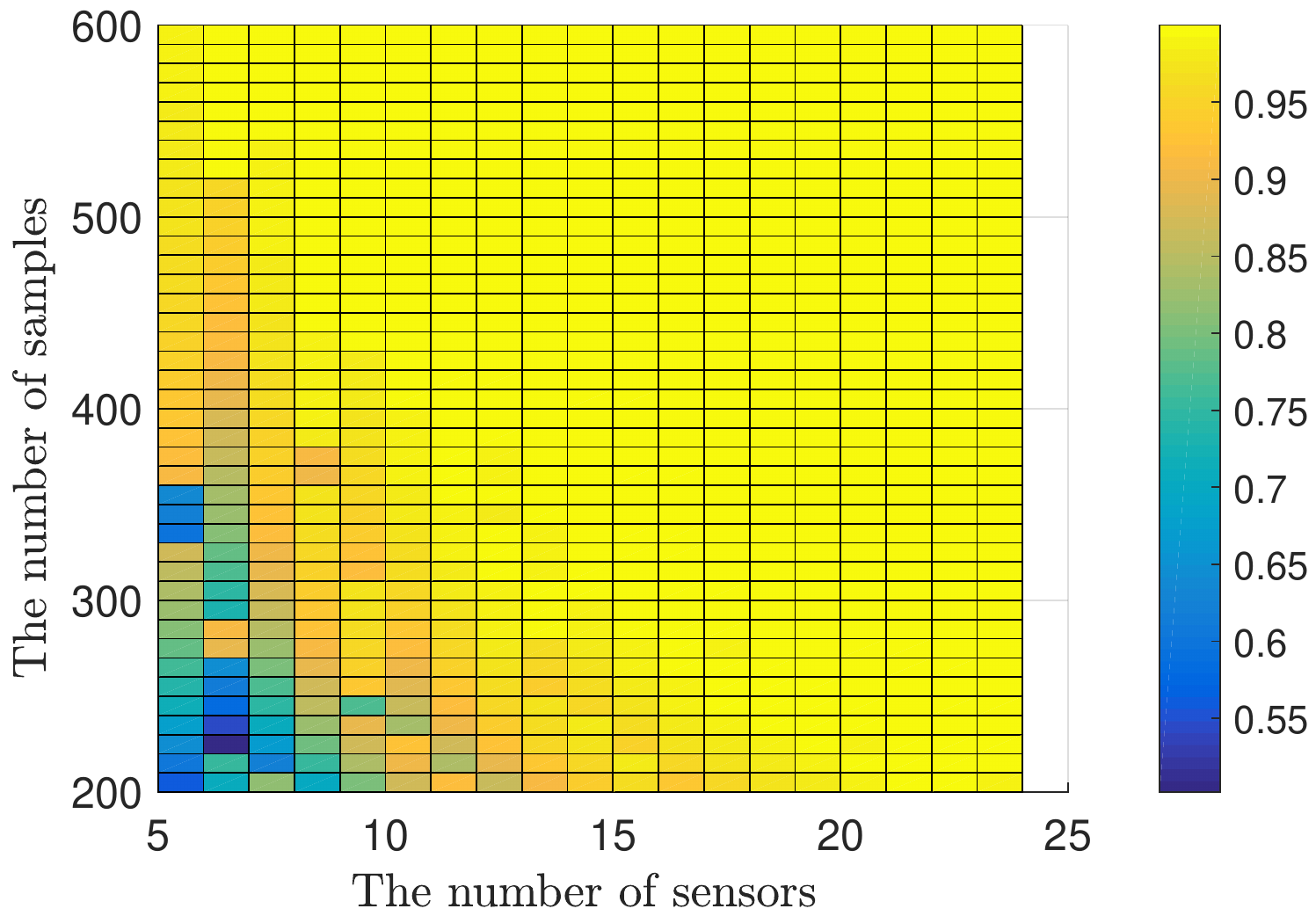}
\caption{The impact of the number of sensors and the number of samples on the drone detection and classification performance.}
\label{Fig-sample-nu-global1}
\end{figure*}

To further improve the drone detection and classification performance, in the simulation results of Fig. \ref{Fig-sample-nu-global1}, we study the impact of the number of sensors and the number of samples for the given false alarm constraints. It is shown that as the number of samples and the number of sensors increase, the global detection probability increases. It means that we can obtain better drone detection and classification performance at the cost of an increasing of the detection delay and the coordination of multiple sensors. Interestingly, the marginal gain is minor as either the number of sensors or the number of samples is sufficiently large.

%

\section{Conclusion}
This article developed an amateur drone surveillance system based on cognitive IoT. We firstly presented a brief survey on the state-of-the-art studies on anti-drone techniques. Then, we proposed a vision named \emph{Dragnet} by tailoring the recent emerging cognitive IoT framework for amateur drone surveillance. Next, the key enabling techniques accompanied with the technical challenges and open issues ahead were discussed. Furthermore, an exemplary simulation on the detection of illegal amateur drone was provided. We firmly believe this important area will be a fruitful research direction and we have just touched one tip of the iceberg. We hope this article will stimulate much more research interests.


\newpage

\textbf{Guoru Ding} (dr.guoru.ding@ieee.org) is an assistant professor in College of Communications Engineering and a Postdoctoral Research Associate at the National Mobile Communications Research Laboratory, Southeast University, Nanjing, China. He received his B.S. degree from Xidian University in 2008 and his PhD degree from the College of Communications Engineering, Nanjing, China, in 2014. His research interests include cognitive radio networks, massive MIMO, machine learning, and big data analytics over wireless networks.

\textbf{Qihui Wu} (wuqihui2014@sina.com) is a professor in the College of Electronic and Information Engineering, Nanjing University of Aeronautics and Astronautics, Nanjing, China. From 2005 to 2007, he was an Associate Professor with the College of Communications Engineering, PLA University of Science and Technology, Nanjing, China, where he served as a Professor from 2008 to 2016. From March 2011 to September 2011, he was an Advanced Visiting Scholar in Stevens Institute of Technology, Hoboken, USA.

\textbf{Linyuan Zhang} (zhanglinyuan5@163.com) received his B.S. degree (with honors) in electronic engineering from Inner Mongolia University, Hohhot, China, in 2012 and his M.S. degree in communications and information system at College of Communications Engineering, PLA University of Science and Technology, in 2015. He is currently pursuing his PhD degree at College of Communications Engineering, PLA University of Science and Technology. His research interests include wireless communications and cognitive radio networks.

\textbf{Yun Lin} (linyun@hrbeu.edu.cn) is an associate professor at college of information and communication engineering, Harbin Engineering University, China. He received the B.S. degree in Dalian Maritime University in 2003, the M.S. degree in Harbin Institute of Technology in 2005, and the Ph.D degree in Harbin engineering university in 2010. He was a visiting scholar in Wright State University, USA, from 2014 to 2015. His research interests include machine learning, information fusion, cognitive and software defined radio.

\textbf{Theodoros A. Tsiftsis} (theodoros.tsiftsis@nu.edu.kz) is an Associate Professor of communication technologies with the School of Engineering, Nazarbayev University, Astana, Kazakhstan. His research interests include the broad areas of cooperative communications, cognitive radio, communication theory, wireless powered communication systems, and optical wireless communication systems. He is currently an Area Editor for Wireless Communications II of the IEEE TRANSACTIONS ON COMMUNICATIONS and an Associate Editor of the IEEE TRANSACTIONS ON MOBILE COMPUTING.

\textbf{Yu-Dong Yao} (yyao@stevens.edu) has been with Stevens Institute of Technology, Hoboken, New Jersey, since 2000, and is currently a professor and department director of electrical and computer engineering. He is an IEEE Fellow, a Fellow of National Academy of Inventors and a Fellow of the Canadian Academy of Engineering. His research interests include wireless communications and networks, spread spectrum and CDMA, antenna arrays and beamforming, cognitive and software-defined radio, and digital signal processing for wireless systems.

\end{document}